\documentclass{PoS}
\usepackage{epsfig}
\usepackage{bm}

\PoS{PoS(LAT2005)261}

\title{Lattice simulations of Born-Infeld non-linear QED }

\ShortTitle{Lattice simulations of Born-Infeld non-linear QED }

\author{\speaker{D.~K.~Sinclair}\thanks{This work was supported by the U.S. 
Department of Energy, Division of High Energy Physics, Contract \newline
W-31-109-ENG-38.}\\
HEP Division, Argonne National Laboratory, 9700 South Cass Avenue, Argonne, 
IL 60439, USA\\
        E-mail: \email{dks@hep.anl.gov}}

\author{J.~B.~Kogut\thanks{Supported in part by NSF grant NSF PHY03-04252.}\\
Department of Energy, Division of High Energy Physics, Washington, DC 20585, 
USA\\
and\\
Dept. of Physics -- TQHN, Univ. of Maryland, 82 Regents Dr., College
Park, MD 20742, USA\\
        E-mail: \email{jbkogut@umd.edu}}

\abstract{Born-Infeld non-linear electrodynamics was introduced to render the
self energy of a point particle finite. It has recently been revived as a field
theory for branes and strings. We quantize this theory on a Euclidean
space-time lattice, using Metropolis Monte-Carlo simulations to measure the
properties of the quantum field theory. L\"{u}scher-Weisz methods are used to
measure the electromagnetic fields from a static point charge. The {\bf D}
field from a point charge appears to be identical to that for the normal
Maxwell Lagrangian. The {\bf E} field is enhanced by quantum fluctuations, and
shows short distance screening as it does in the classical theory.}

\FullConference{XXIIIrd International Symposium on Lattice Field Theory\\
		 25-30 July 2005\\
		 Trinity College, Dublin, Ireland}

\begin{document}

\section{Introduction}

The $n+1$ dimensional Born-Infeld (non-linear electrodynamics) action 
\cite{Born:1934,Born:1934gh} is:
\begin{equation}
S = b^2 \int d^{n+1}x\,\left[1-\sqrt{-\det\left(g_{\mu\nu}+
\frac{1}{b}F_{\mu\nu}\right)}\,\right].
\label{eqn:Sminkowski}
\end{equation}
This has seen a revival as a theory of strings and branes 
\cite{Fradkin:1985qd,Leigh:1989jq,Aganagic:1996nn,Gibbons:1997xz,
Tseytlin:1999dj}.
Choosing $n=9$, and dimensionally reducing this action from $9+1$ to $p+1$ 
dimensions describes a $p$-brane. The $9-p$ additional components of $A_\mu$ 
are identified with the transverse components of the string/brane.

Most of the serious work on these theories has dealt with their classical 
behaviour \cite{Callan:1997kz}.  
We are simulating these quantum theories on the lattice. We are 
starting with the simplest case where $n=p=3$, the original Born-Infeld 
modification of electrodynamics, designed to make the self energy of a point
charge finite.

Section~2 reviews the classical Born-Infeld theory. In section~3 we indicate
how this is ported to the lattice allowing Monte-Carlo simulations of the 
quantum theory. Section~4 details our simulations and preliminary results.
A discussion of our results and conclusions are given in section~5.

\section{Classical Born-Infeld electrodynamics in Minkowski space-time}

This section summarises those results in 
references~\cite{Bialynicki-Birula:1984tx,Chruscinski:1997pe,Chruscinski:1997jd,
Chruscinski:1998uw}
which are relevant for our investigations.

Evaluating the determinant in equation~\ref{eqn:Sminkowski} the Lagrangian in  
$3+1$~dimensions is
\begin{equation}
{\cal L} = b^2 [1-\sqrt{1-b^{-2}(\bm{E}^2-\bm{B}^2) - b^{-4}(\bm{E\cdot B})^2}].
\end{equation}
One can now define $\bm{D}$ and $\bm{H}$ by
\begin{eqnarray}
\bm{D} &=& {\partial{\cal L} \over \partial\bm{E}} =
               {\bm{E}+b^{-2}(\bm{E\cdot B})\bm{B} \over
               \sqrt{1-b^{-2}(\bm{E}^2-\bm{B}^2)-b^{-4}(\bm{E\cdot B})^2}}  
\nonumber \\
\bm{H} &=& {\partial{\cal L} \over \partial\bm{B}} =
               {\bm{B}-b^{-2}(\bm{E\cdot B})\bm{E} \over
               \sqrt{1-b^{-2}(\bm{E}^2-\bm{B}^2)-b^{-4}(\bm{E\cdot B})^2}}.
\label{eqn:DH}
\end{eqnarray}
Interaction with charged particles is implemented, as usual, by adding a term
$j_\mu A^\mu$ to the Lagrangian. In terms of $\bm{E}$, $\bm{B}$, $\bm{D}$ and 
$\bm{H}$, the equations of motion are the standard Maxwell equations. The 
non-linearity is hidden in equations~\ref{eqn:DH}.
     
For a static point charge $\rho = e \delta^3(\bm{r})$ the electric fields are
\begin{eqnarray}
\bm{D} & = & {e \over 4\pi r^2} \bm{\hat{r}}    \nonumber     \\
\bm{E} & = & {e \over 4\pi}{\bm{\hat{r}} \over \sqrt{r^4 + r_0^4}},
\end{eqnarray} 
where $\bm{\hat{r}} = {\bm{r} \over r}$ and $r_0 = \sqrt{|e| \over 4\pi b}$.
Hence the $\bm{D}$ field for a static point charge is identical to the Maxwell
solution, while the $\bm{E}$ field is screened at short distances.

\section{Lattice Born-Infeld quantum-electrodynamics}

The Euclidean space action for Born-Infeld QED
\begin{equation}
S = b^2 \int d^4 x [\sqrt{1 + b^{-2}(\bm{E}^2+\bm{B}^2) 
  + b^{-4}(\bm{E\cdot B})^2}-1]
\end{equation}
is positive. Hence it can be simulated using Monte-Carlo methods. 

On the lattice we use the non-compact formulation:
\begin{equation}
F_{\mu\nu}(x+{\textstyle \frac{1}{2}}\hat{\mu}
            +{\textstyle \frac{1}{2}}\hat{\nu})
= A_\nu(x+\hat{\mu})-A_\nu(x)-A_\mu(x+\hat{\nu})+A_\mu(x)
\end{equation}
and average over the 16 choices of 6 plaquettes associated with each lattice
site. We also define $\beta=b^2 a^4$ where $a$ is the lattice spacing.
Simulations are performed using the Metropolis Monte-Carlo method
\cite{Metropolis:1953am}.

We measure the $\bm{E}$ and $\bm{D}$ fields due to a static point charge. This
point charge $e$ is introduced by including a Wilson Line (Polyakov Loop)
$W(\bm{x})$.
\begin{equation}
W(\bm{x}) = \exp\left\{i e \sum_t \left[A_4(\bm{x},t) 
- \frac{1}{N_x N_y N_z}\sum_{\bm{y}} A_4(\bm{y},t)\right]\right\}
\end{equation}
The second (`Jellium') term is needed, since a net charge would be 
inconsistent with periodic boundary conditions on $A_\mu$. 
$\langle\bm{E}\rangle$ and $\langle\bm{D}\rangle$ in the presence of this 
charge are given by
\begin{eqnarray}
i \langle\bm{E}\rangle_\rho (\bm{y}-\bm{x}) &=&
{\langle\bm{E}(\bm{y},t)W(\bm{x})\rangle \over \langle W(\bm{x})\rangle}   
\nonumber \\
i \langle\bm{D}\rangle_\rho (\bm{y}-\bm{x}) &=&
{\langle\bm{D}(\bm{y},t)W(\bm{x})\rangle \over \langle W(\bm{x})\rangle}.
\end{eqnarray}
Since $W$ is complex, there is a sign problem, which causes 
$\langle W(\bm{x})\rangle$ to fall exponentially with $N_t$. We use the method
of L\"{u}scher and Weisz \cite{Luscher:2001up} 
(Parisi, Petronzio and Rapuano \cite{Parisi:1983hm})  
with thickness 1 and 2
timeslices to overcome this exponential factor.

\section{Simulations and Results}

\begin{figure}[bht]
\begin{minipage}[t]{0.48\textwidth}
\epsfig{file=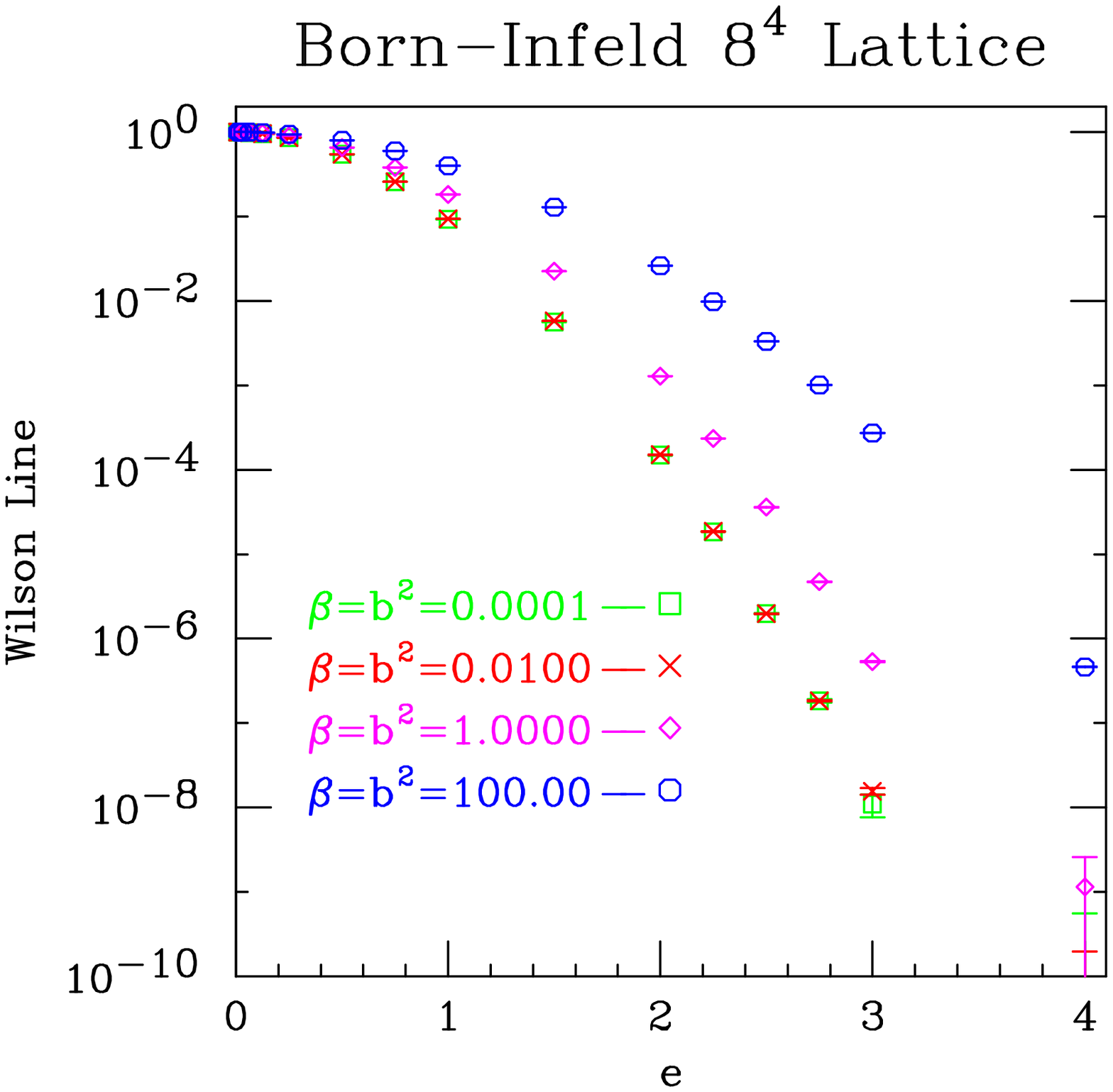, width=\textwidth}
\caption{Wilson Lines as functions of charge $e$, for a
range of $\beta=a^2b^2$.}\label{fig:wilson}
\end{minipage}
\hfill
\begin{minipage}[t]{0.48\textwidth}
\epsfig{file=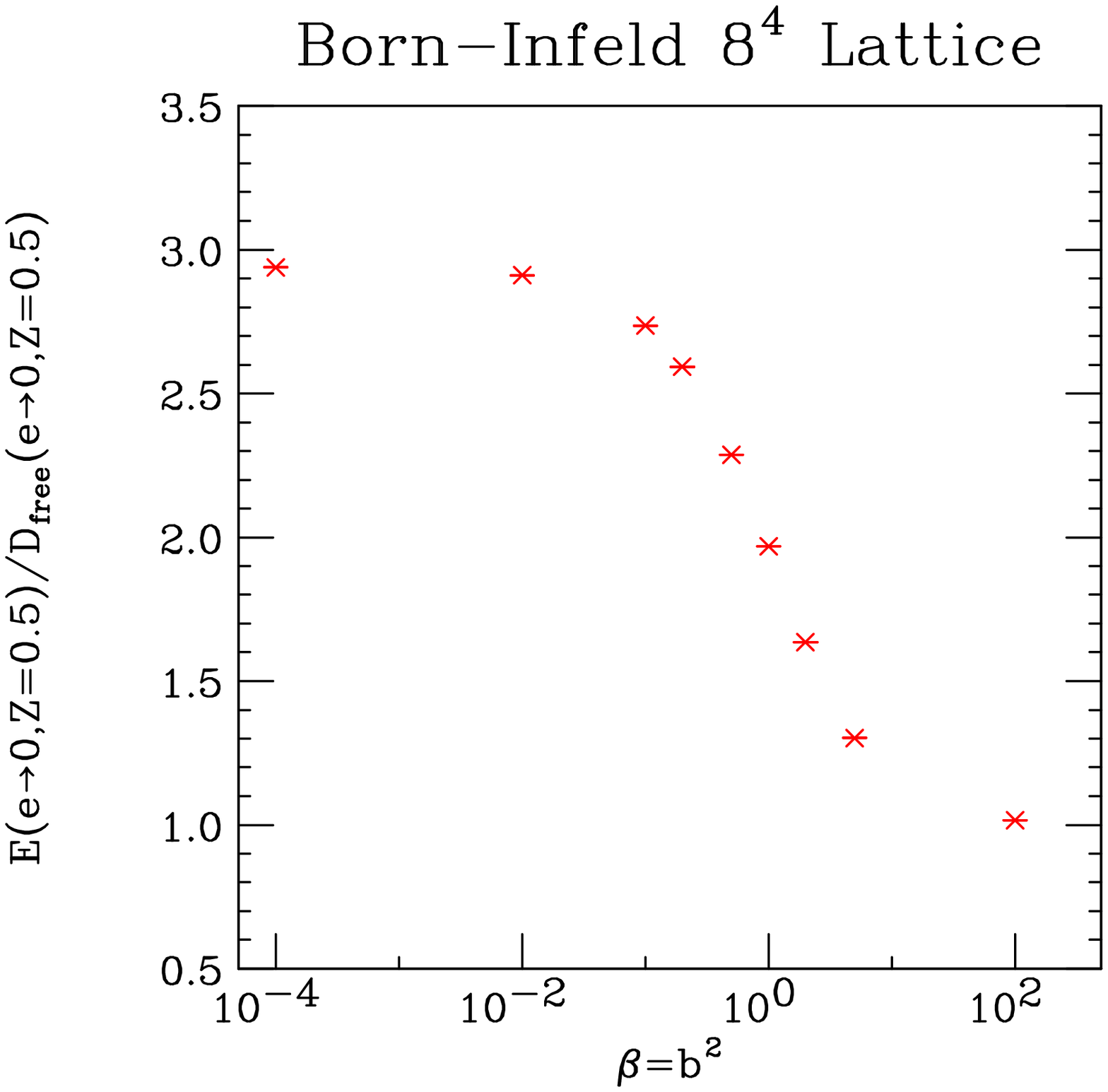,width=\textwidth}
\caption{$E/D$ ratio at minimum separation for $e \rightarrow 0$ as a function
of $\beta$.}\label{fig:e/d}
\end{minipage}
\end{figure}

We have performed preliminary simulations of 500,000 10-hit Metropolis sweeps 
of the lattice at $\beta=100,1.0,0.01,0.0001$ and 100,000 sweeps at 
$\beta=5,2,0.5,0.2,0.1$, making measurements every 100 sweeps. We measured the 
$\bm{E}$ and $\bm{D}$ fields for on axis separations from the point charge.
Figure~\ref{fig:wilson} shows the expectation value of the Wilson lines
(Polyakov loops) obtained from these simulations. Note that the value falls
rapidly with increasing $e$. The rate of falloff also increases with
increasing non-linearity (decreasing $\beta$). Note also the small relative
errors, even when the magnitude has fallen $8$ orders of magnitude, which shows
the effectiveness of the L\"{u}scher-Weisz method.

Figure~\ref{fig:e/d} shows the ratio of the $\bm{E}$ field in the direction of
the separation from the charge to the $\bm{D}=\bm{E}$ field for the free field
(Maxwell) theory, at the minimum separation ($Z=0.5$), in the limit of zero
charge. At large $\beta$, where the Born-Infeld theory asymptotes to the
Maxwell theory, this ratio approaches $1$. Classically, this ratio is $1$ for
all values of $b$. Quantum fluctuations cause this ratio to increase with
increasing non-linearity (decreasing $\beta$), approaching a value close to
$3$ for small $\beta$.

\begin{figure}[tbh]
\epsfig{file=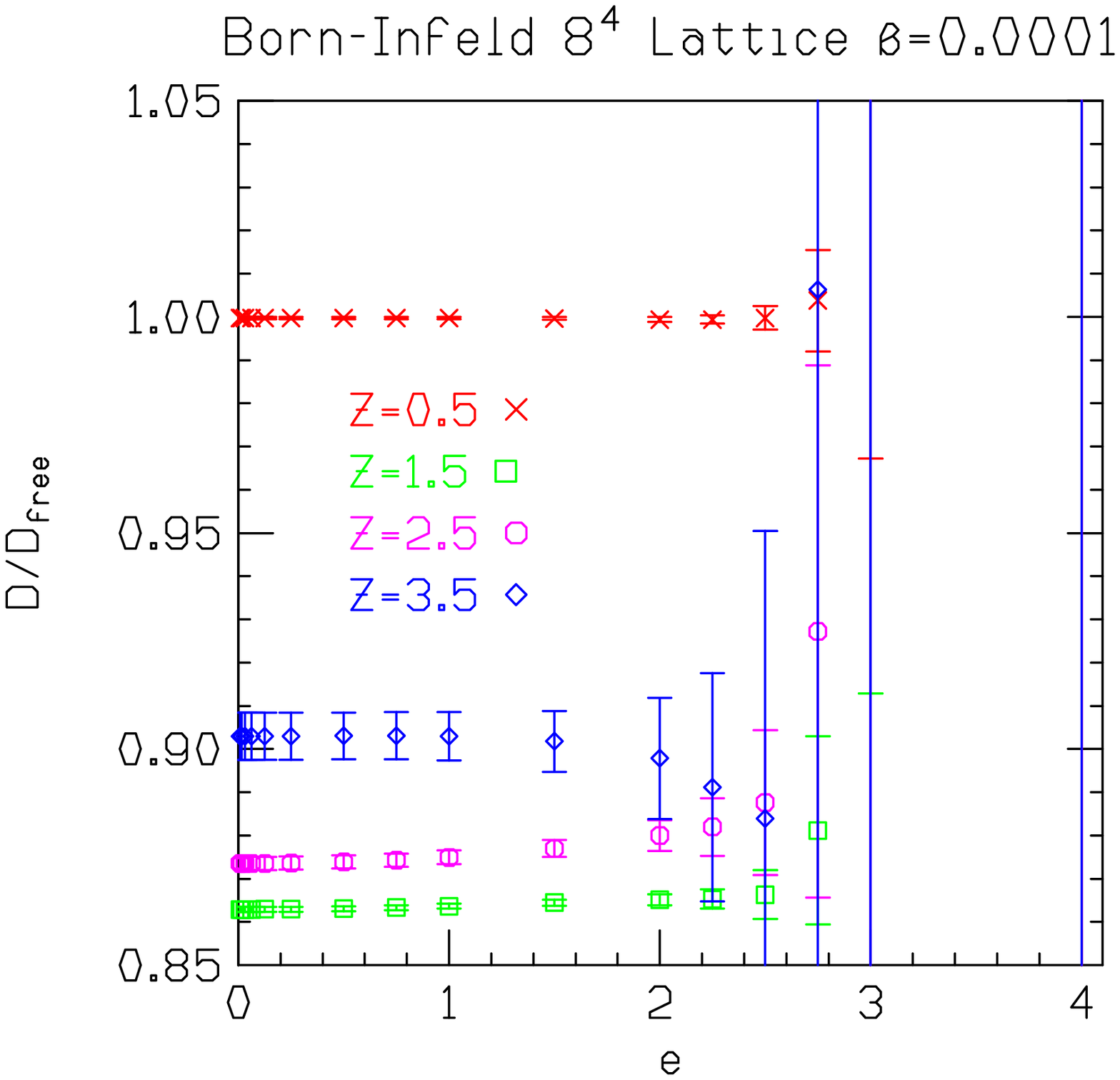,width=0.5\textwidth}
\epsfig{file=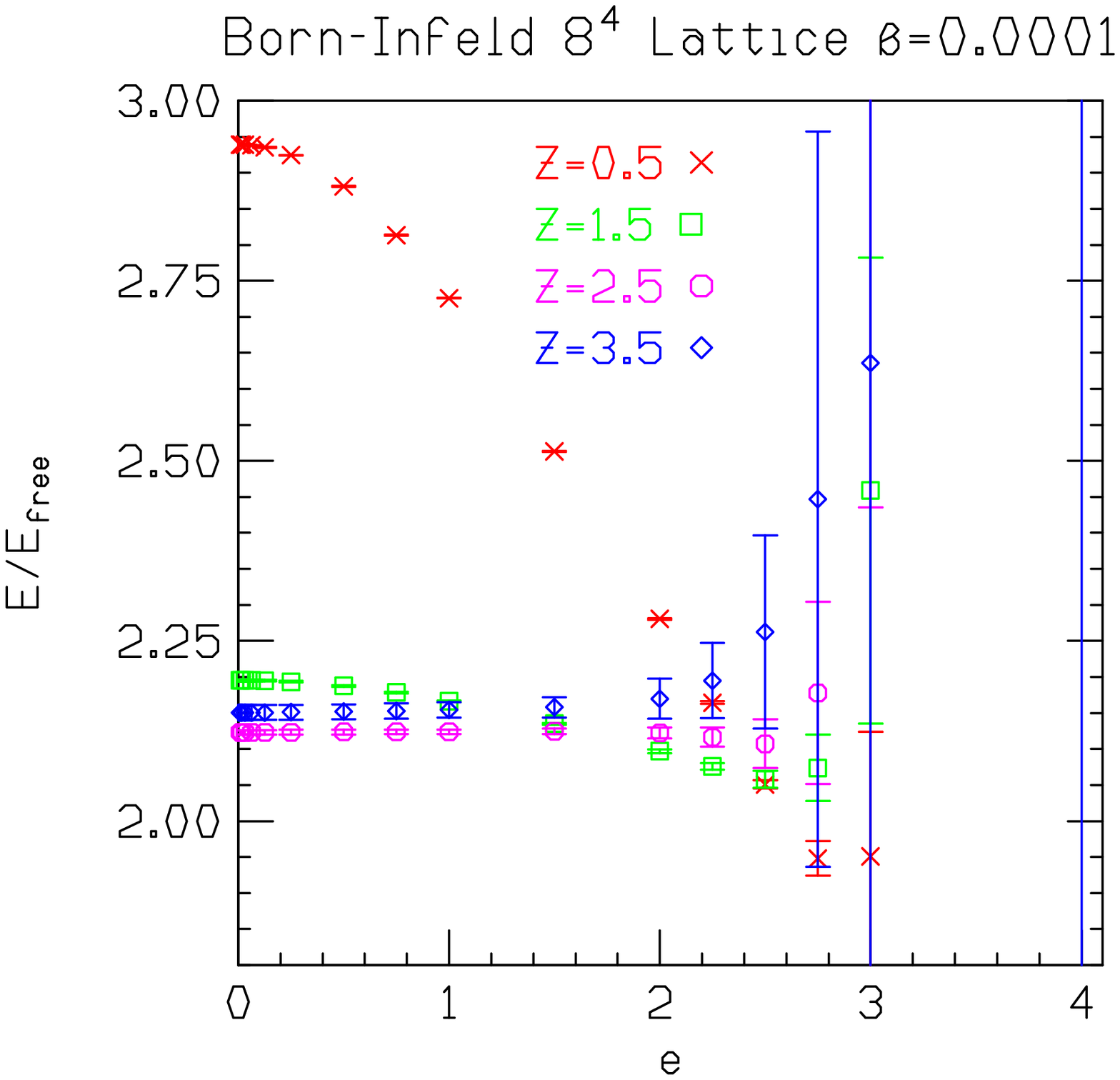,width=0.5\textwidth}
\caption{a) $\bm{D}$ fields at distance $Z$ from the point charge $e$, scaled by
the free field (Maxwell) values, for $\beta=0.0001$.
b) $\bm{E}$ fields at distance $Z$ from the point charge $e$, scaled by
the free field (Maxwell) values, for $\beta=0.0001$.}
\label{fig:e&d}
\end{figure}

In figure~\ref{fig:e&d}a, we plot the $\bm{D}$ fields scaled by their free field
values at each separation, as a function of charge, for $\beta=0.0001$, where
the non-linearity is large, and the $\bm{D}$ field comes almost entirely from
the $(\bm{E\cdot B})\bm{B}$ term in its definition. The fact that this ratio is 
still 1 for all $e$ at minimal separation is because $\bm{D}$ still obeys 
$\bm{\nabla\cdot D}=\rho$, combined with cubic symmetry. Because we do not have
full rotational symmetry on the lattice $\bm{\nabla\cdot D}=\rho$ is
insufficient to make this ratio 1 at other separations. The fact that this
ratio is never more than 15\% from 1 suggests that it would be 1 if we had
rotational symmetry. However, we would expect rotational symmetry to be
restored at large distances, which is why the ratio is closer to 1 for larger
separations. Figure~\ref{fig:e&d}b is a similar graph for the $\bm{E}$ field.
As well as showing the effects of quantum fluctuations as in
figure~\ref{fig:e/d}, the $\bm{E}$ field is clearly screened at short
distances as $e$ increases, similar to what is seen in the classical theory.

What is different from the classical theory is that classically the screening
length continues to increase with decreasing $b$. The quantum theory approaches
a limit as $\beta \rightarrow 0$.

\section{Discussion and conclusions}

We have succeeded in using lattice Monte-Carlo methods to extract 
non-perturbative physics from Born-Infeld electrodynamics, quantized using the
Euclidean-time functional integral approach. The on-axis (quantum) 
electrostatic fields of a point charge are measured as functions of the charge
$e$ introduced as a Wilson Line. The approach of L\"{u}scher and Weisz, which
reduces these measurements from an exponential- to a polynomial-time problem,
was essential for extracting these quantities.

In the classical field-theory $E/D \rightarrow 1$ as $e \rightarrow 0$.
For the quantum theory $E/D$ increases from $1$ as the nonlinearity is 
increased indicating that the dielectric constant $\epsilon < 1$.

As $|e|$ is increased, the $\bm{E}$ field is screened at short distances.
Screening increases with $|e|$ and with increasing nonlinearity ($\beta$ or 
$b$). The screening length $r_0$ appears to increase as $\sqrt{|e|}$ as for 
the classical theory. $\bm{D}$ shows no such screening and appears to
independent of the non-linearity, while $\bm{D}/e$ appears independent of $e$.

Unlike the classical theory, where the screening length diverges as 
$b \rightarrow 0$, the quantum theory approaches a fixed-point field theory as
$\beta=b^2 a^4 \rightarrow 0$. This conformal field theory has Euclidean 
Lagrangian ${\cal L}_E = \frac{1}{4}|\bm{E.B}|$ and Hamiltonian  
${\cal H} = |\bm{D} \times \bm{B}|$ 
\cite{Bialynicki-Birula:1984tx,Chruscinski:2000zm}.

Normally, Born-Infeld QED is considered as an effective field theory with a
momentum cutoff. However, as this cutoff $\rightarrow \infty$ it approaches
the above fixed-point theory. If this fixed-point field theory is non-trivial,
it would serve to define Born-Infeld QED without a cutoff.

These first simulations were performed on $8^4$ lattices. We are extending
these simulations to larger lattices. We then plan to study those $p$-brane
theories obtained by dimensional reduction of $n+1$ dimensional Born-Infeld
theories to determine if the quantized theories continue to show string/brane
dynamics.

\section*{Acknowledgements}
Our simulations are currently running on the Rachael supercomputer at the
Pittsburgh Supercomputer Center.

\end{document}